\documentclass[aps,prb,amsmath,amssymb,twocolumn,superscriptaddress,floatfix]{revtex4-2}
\usepackage{graphicx}
\usepackage{hyperref}
\usepackage{bm}
\usepackage[dvipsnames]{xcolor}
\hypersetup{pdftitle={Needles},pdfauthor={Darren},pdfsubject={Frustrated magnetism},pdfdisplaydoctitle}

\def\rouaH{Cu$_\text{2}$(OH)$_\text{3}$NO$_\text{3}$}

\def\botHBr{Cu$_\text{2}$(OH)$_\text{3}$Br}

\def\brochH{Cu$_\text{4}$SO$_\text{4}$(OH)$_\text{6}$}

\def\cue{$\mathbf{q}$}

\def\TN{\ensuremath{T_\text{N}}}

\def\cmag{\ensuremath{c_\text{mag}}}
\begin{document}

\title{Stacking disorder in novel ABAC-stacked brochantite, \brochH}

\author{Aswathi Mannathanath Chakkingal}
\affiliation{Institut f\"ur Festk\"orper- und Materialphysik, Technische Universit\"at Dresden, 01062 Dresden, Germany}

\author{Chloe Fuller}
\affiliation{ESRF, The European Synchrotron, 71 Avenue des Martyrs, CS 40220, 38043 Grenoble Cedex 9, France}

\author{Maxim Avdeev}
\affiliation{Australian Nuclear Science and Technology Organisation, Lucas Heights, NSW 2234, Australia}
\affiliation{School of Chemistry, The University of Sydney, Sydney, NSW 2006, Australia}

\author{Roman Gumeniuk}
\affiliation{Institut für Experimentelle Physik, TU Bergakademie Freiberg, 09596 Freiberg, Germany}

\author{Kaushick K.\ Parui}
\affiliation{Institut f\"ur Festk\"orper- und Materialphysik, Technische Universit\"at Dresden, 01062 Dresden, Germany}

\author{Marein C.\ Rahn}
\altaffiliation[Current affiliation:~]{Institut f\"ur Physik, Universit\"at Augsburg, 86159 Augsburg, Germany}
\affiliation{Institut f\"ur Festk\"orper- und Materialphysik, Technische Universit\"at Dresden, 01062 Dresden, Germany}

\author{Falk Pabst}
\author{Yiran Wang}
\affiliation{Professur f\"ur Anorganishe Chemie II, Technische Universit\"at Dresden, 01062 Dresden, Germany}

\author{Sergey Granovsky}
\affiliation{Institut f\"ur Festk\"orper- und Materialphysik, Technische Universit\"at Dresden, 01062 Dresden, Germany}

\author{Artem Korshunov}
\author{Dmitry Chernyshov}
\affiliation{ESRF, The European Synchrotron, 71 Avenue des Martyrs, CS 40220, 38043 Grenoble Cedex 9, France}

\author{Dmytro S.\ Inosov}
\email{dmytro.inosov@tu-dresden.de}
\affiliation{Institut f\"ur Festk\"orper- und Materialphysik, Technische Universit\"at Dresden, 01062 Dresden, Germany}
\affiliation{W\"urzburg-Dresden Cluster of Excellence on Complexity and Topology in Quantum Matter\,---\,ct.qmat, Technische Universit\"at Dresden, 01062 Dresden, Germany}

\author{Darren C.\ Peets}
\email{darren.peets@tu-dresden.de}
\affiliation{Institut f\"ur Festk\"orper- und Materialphysik, Technische Universit\"at Dresden, 01062 Dresden, Germany}

\begin{abstract}

In geometrically frustrated magnetic systems, weak interactions or slight changes to the structure can tip the delicate balance of exchange interactions, sending the system into a different ground state.  Brochantite, \brochH, has a copper sublattice composed of distorted triangles, making it a likely host for frustrated magnetism, but exhibits stacking disorder.  The lack of synthetic single crystals has limited research on the magnetism in brochantite to powders and natural mineral crystals.  We grew crystals which we find to be a new polytype with a tendency toward ABAC stacking and some anion disorder, alongside the expected stacking disorder.  Comparison to previous results on natural mineral specimens suggests that cation disorder is more deleterious to the magnetism than anion and stacking disorder.  Our specific heat data suggest a double transition on cooling into the magnetically ordered state.

\end{abstract}
\maketitle 

\section{Introduction}
In magnetically frustrated systems, the leading exchange interactions compete with each other, and the magnetic ground state can depend crucially on weaker interactions, or on details of the crystal structure.  Such systems can be exquisitely tunable, since minor perturbations can upset this delicate balance and send the system into a completely different magnetically ordered state.  This situation is most commonly realized by arranging the magnetic ions in a geometry that pits different interactions against each other\,\cite{Ramirez1994,Lacroix2011,Diep2013,Batista2016,Schmidt2017}, with the classic example being spins on a triangle with antiferromagnetic nearest-neighbour interactions.  The competition among interactions that destabilizes conventional forms of magnetic order can be aided by limiting the number of interactions at each site, for instance in low-dimensional systems, or by quantum fluctuations.  The latter are most relevant for small spins, especially $S=1/2$, as is the case for Cu$^{2+}$.

\begin{figure}[tb]
  \includegraphics[width=\columnwidth]{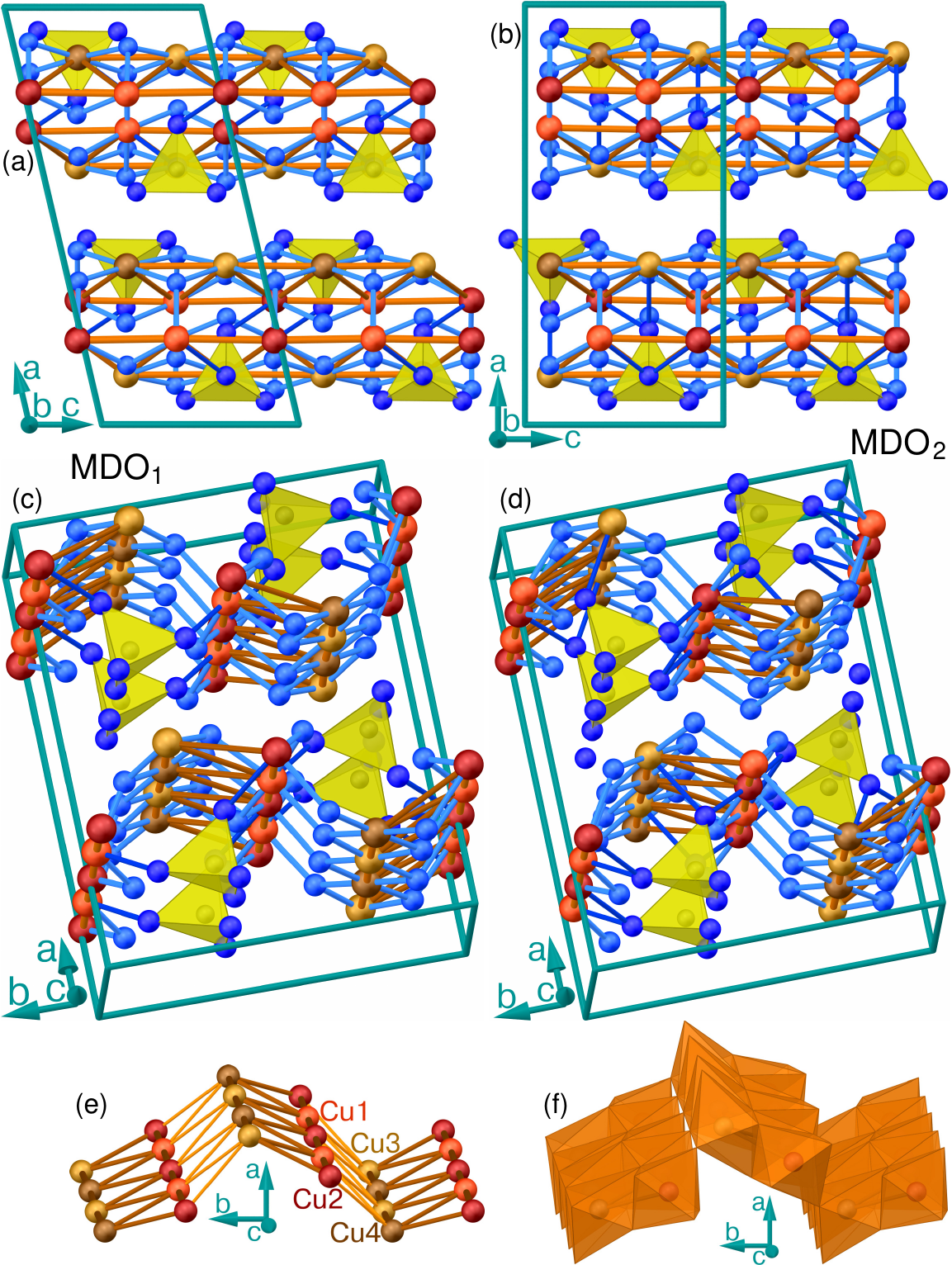}
  \caption{\label{MDO}Crystal structure of brochantite: 
 The (a) MDO$_1$ and (b) MDO$_2$ polytypes, based on Ref.~\onlinecite{Merlino2003}, viewed along $b$ to show the different stacking patterns.  (c,d)  The same polytypes, showing the layered structure.  Panels (e) and (f) highlight a single Cu layer, represented in terms of Cu--Cu linkages and CuO$_6$ polyhedra, respectively.  Oxygen atoms are shown in blue, SO$_4^{2-}$ tetrahedra are yellow, and the four Cu sites are labelled in panel (e); hydrogen positions were not refined in this reference.}
\end{figure}

The stability of the $3d^9$ electronic configuration of Cu$^{2+}$ makes it a particularly accessible magnetic ion for quantum magnetism, while its low spin-orbit coupling makes it a nearly pure-spin moment.  The copper-based minerals are of particular interest from a magnetic frustration standpoint, since their copper sublattices are typically composed of distorted Cu$^{2+}$ triangles\,\cite{Inosov2018}, and have proven a rich platform for novel physics.  Examples include the candidate quantum spin-liquid state in herbertsmithite ZnCu$_3$(OH)$_6$Cl$_2$\,\cite{Shores2005,Norman2016,Lancaster2023}; enormous effective moments in atacamite Cu$_2$Cl(OH)$_3$\,\cite{Heinze2021}; misfit multiple-\cue\ order in antlerite Cu$_3$SO$_4$(OH)$_4$\,\cite{Kulbakov2022a,Kulbakov2022b}; possible spinon-magnon interactions in botallackite \botHBr\,\cite{Zhang2020};  and a helically modulated cycloidal state in rouaite \rouaH\,\cite{Aswathi2024}.

The copper sublattice in brochantite, \brochH, is composed of rippled Cu$^{2+}$ planes comprising quasi-one-dimensional (1D) two-leg triangular ladders\,\cite{Cocco1959,Helliwell1997,Merlino2003}.  Copper oxide polyhedra are assembled edge-sharing within the ladders while adjacent ladders within a plane are corner sharing.  Linkages between planes are through sulphate groups and hydrogen bonds.  Inelastic neutron scattering on natural mineral samples confirms that the exchange interactions are quite one-dimensional\,\cite{Nikitin2023}.  The stacking of these rippled layers is known to be complex and partially disordered, with multiple polytypes reported\,\cite{Merlino2003}, and even an orthorhombic variant ``orthobrochantite'' with a doubled unit cell which was approved in 1978 but never formally published --- this has since been discredited as the MDO$_1$ polytype\,\cite{Mills2010}.  The distinction between ``maximum degree of order'' polytypes MDO$_1$ and MDO$_2$ (Fig.~\ref{MDO}) centres on the shift along $c$ between consecutive Cu layers, which leads to a significant monoclinic angle in MDO$_1$.  These two forms of stacking are nearly degenerate in energy, leading to the complex stacking behaviour and disorder.  While the stacking direction is expected to have the weakest exchange interactions, in both antlerite and rouaite we found that weak interactions between the Cu ladders or layers lead to helical order, so the stacking may still play a significant role.  Brochantite thus offers an interesting platform for studying frustrated low-dimensional physics in the presence of stacking disorder or multiple stacking arrangements.  However, studying this comprehensively requires chemical control over the degree of stacking disorder.  

While the preparation of brochantite powder is relatively straightforward, to date no growth of single crystals has been reported.  This is particularly problematic for research with neutrons, where protons produce an enormous incoherent background from which the weak $S=\frac{1}{2}$ Cu$^{2+}$ magnetism must be distinguished.  Synthetic crystals can be deuterated, but the deuteration level of natural mineral samples is uniformly low.  Natural mineral samples have been studied nonetheless\,\cite{Nikitin2023}, but the ability to prepare deuterated single crystals would make research on this material significantly easier.  Here we report the synthesis of single-crystalline brochantite, present its magnetic behavior, and discuss its stacking disorder.  This represents a synthetic foothold for brochantite crystal growth, and future optimization of growth parameters may lead to some degree of control over the stacking pattern.  The crystals showed a tendency toward ABAC stacking, which would be a new polytype, likely also present in mineral samples.  


\section{Experimental}

The single crystals studied here were grown hydrothermally in a Teflon-lined stainless-steel autoclave.  Cu(OH)$_2$ (Alfa Aesar, 94\%), Al$_2$(SO$_4)_3\cdot16$H$_2$O (Fisher Scientific, $\geq$94\%), NaCl (Fisher Scientific, $\geq$99.5\%), and Li(OH)$\cdot$H$_2$O (Thermo Scientific, 99+\%) were added in the molar ratio 14:1:2:3.4 and well mixed in distilled water.  After five days at 180$^\circ$C, thin, flat needle-shaped crystals of size up to $20\times0.7\times0.01$\,mm$^3$ were obtained, along with CuO powder and aluminum salts.  The final pH of the colourless supernatant was 5--6.  Decanting off this remaining liquid and washing the precipitates with distilled water removed the Na and Li.  The elemental composition of the single-crystalline samples was analyzed using energy-dispersive x-ray (EDX) spectroscopy collected with an Oxford Instruments X-Max Silicon Drift Detector on a Hitachi SU8020 scanning electron microscope.  A 10-nm-thick gold film was sputtered on the crystals before the measurement to ensure good surface conductivity.  We found a homogeneous Cu:S ratio of 3.32(17):1, no impurity phases, and no traces of Al down to the 3\,ppm level.  Chlorine was also not detected.  Note that this stoichiometry appears to be somewhat SO$_4^{2-}$-rich relative to the accepted composition of brochantite. A handful of attempts were made to grow brochantite without Al present in the source materials, but as in previous brochantite syntheses\,\cite{Vilminot2006} these did not produce macroscopic crystals.

Single-crystal laboratory x-ray diffraction data were collected on a Bruker-AXS KAPPA APEX II CCD diffractometer with graphite-monochromated Mo-\(K_{\alpha}\) radiation. Weighted full-matrix least-squares refinements on $F^2$ were performed with {\sc Shelx}\,\cite{Shelx2008,Shelx2015} as implemented in {\sc WinGx} 2014.1\,\cite{WinGX}.  High-resolution synchrotron powder diffraction and single-crystal Bragg and diffuse x-ray diffraction data were collected at room temperature on beamline BM01 at the European Synchrotron Radiation Facility (ESRF), Grenoble, France\,\cite{SNBL2016}, with wavelengths of 0.65524\,\AA\ (single crystals) and 0.68925\,\AA\ (powder).  The powder was prepared by grinding several single crystals. Additional diffuse-scattering data were collected at the diffraction side station of the ID28 beamline at the ESRF\,\cite{ID28}.  These diffuse data were symmetrized by mirroring them about $H=0$, $K=0$, and/or $L=0$, assuming orthorhombic symmetry.  Powder data were refined using the {\sc FullProf}\,\cite{FullProf} software suite; single-crystal synchrotron diffraction data were processed with {\sc CrysAlis} software, then structural analysis was done in {\sc Shelx}.  Structural disorder and diffuse scattering were analyzed at the ESRF with a Monte-Carlo approach using locally written scripts.  

Neutron Laue diffraction patterns of a $\sim$2\,mm-long single crystal were measured for several distinct sample orientations with respect to the incident neutron beam at room temperature using the Koala white-beam neutron Laue diffractometer\,\cite{Koala} at the OPAL Research Reactor, Australian Centre for Neutron Scattering (ACNS), Australian Nuclear Science and Technology Organisation (ANSTO), in Sydney, Australia. Image data processing, including indexing, intensity integration, and wavelength distribution normalization, was performed using LaueG\,\cite{LaueG}.  Crystal structure refinements of the Koala data were carried out using {\sc Jana2020}\,\cite{Jana2020}.

Temperature-dependent magnetization measurements were performed by vibrating sample magnetometry (VSM) in a Cryogenic Ltd.\ Cryogen-Free Measurement System (CFMS), under zero-field-cooled and field-cooled conditions.  Four-quadrant $M$--$H$ loops were also measured at several temperatures.  The single crystals were mounted to a plastic bar using GE varnish for $H\parallel b$ and $c$.  For $H\parallel [100]$ (i.e., $\text{\itshape a}^\text{*}$), the crystals were attached to an acetate film with GE varnish, which was inserted into a plastic straw;  attempts to measure and subtract the acetate contribution were not successful.

Low-temperature specific heat measurements were performed on several single crystals using a Physical Property Measurement System (PPMS) DynaCool-12 from Quantum Design, equipped with a $^3$He refrigerator.  Measurements were taken using both $^3$He and $^4$He specific heat pucks.  Contributions from the sample holder and Apiezon N grease were subtracted.  Multiple data points were collected at each temperature and averaged;  the first data point at each temperature was discarded to exclude the possibility of incomplete thermal stabilization.  

\section{Crystals}

\begin{figure}[tb]
\includegraphics[width=\columnwidth]{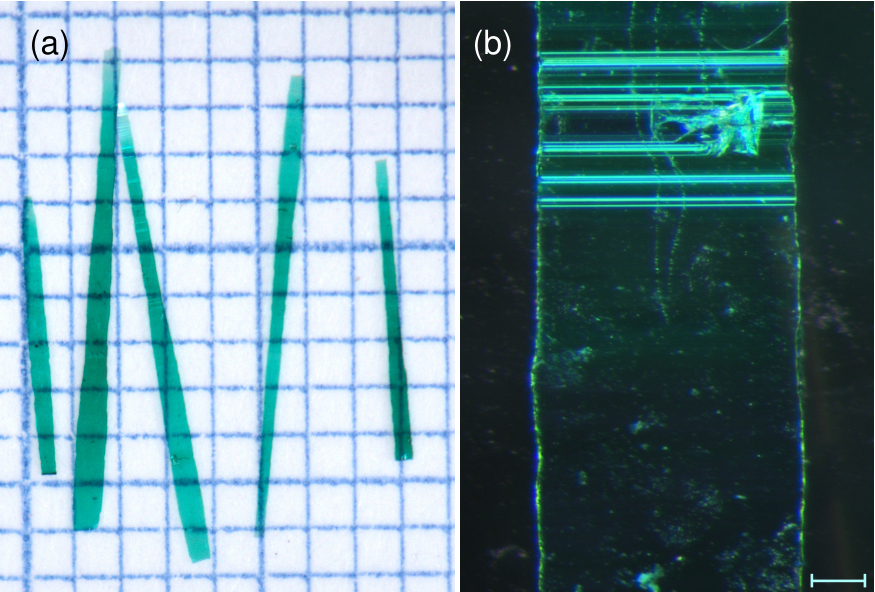}
\caption{\label{photo}(a) Several characteristic crystals on mm-ruled graph paper.  (b) Detail of a defect region in a crystal by optical microscopy (transmission) with crossed polarizers;  the scale bar is 0.1\,mm.  There were no other such defects in this crystal.}
\end{figure}

Crystals formed as thin, flat needles, up to 20\,mm long, up to 1\,mm wide, and approximately 10\,$\mu$m thick --- several representative crystals are shown in Fig.~\ref{photo}(a).  As may be expected, these were quite fragile.  The crystals formed with large (100) faces, and the long direction was $b$.  Throughout this paper we use the unit cell orientation of the MDO polytypes (Fig.~\ref{MDO}), and we use [100] $(\text{\itshape a}^\text{*})$ rather than $a$ even in the orthorhombic setting since [100] is the same in both polytypes.  An optical microscopy check for obvious signs of twinning with crossed polarizers in transmission mode found features only around large defects, as shown in Fig.~\ref{photo}(b), which were present in only a few of the crystals.  Ridges visible in the reflection from a few of the crystals in Fig.~\ref{photo}(a) are associated only with variations in thickness.  

The yield of this growth approach is relatively low, with the majority of the precursors forming CuO and simple Cu and Al salts.  Perhaps as a consequence of this, no intergrowth of crystals was observed.  It remains unclear why Al$_2$(SO$_4$)$_3$ seems to be important for obtaining crystals and whether Cl$^-$ plays a role.  The excess of sulfate may play a role, the Al$^{3+}$ may be involved in the growth, or we may be altering how the cations are solvated, for example.  It is likely that an understanding of this issue would lead to improvements in yield, and possibly to control over the polytype grown.  Our growth of brochantite single crystals, while not fully optimized, represents an important foothold which will enable future optimization. 

\section{Specific Heat}

\begin{figure}[tb]
  \includegraphics[width=\columnwidth]{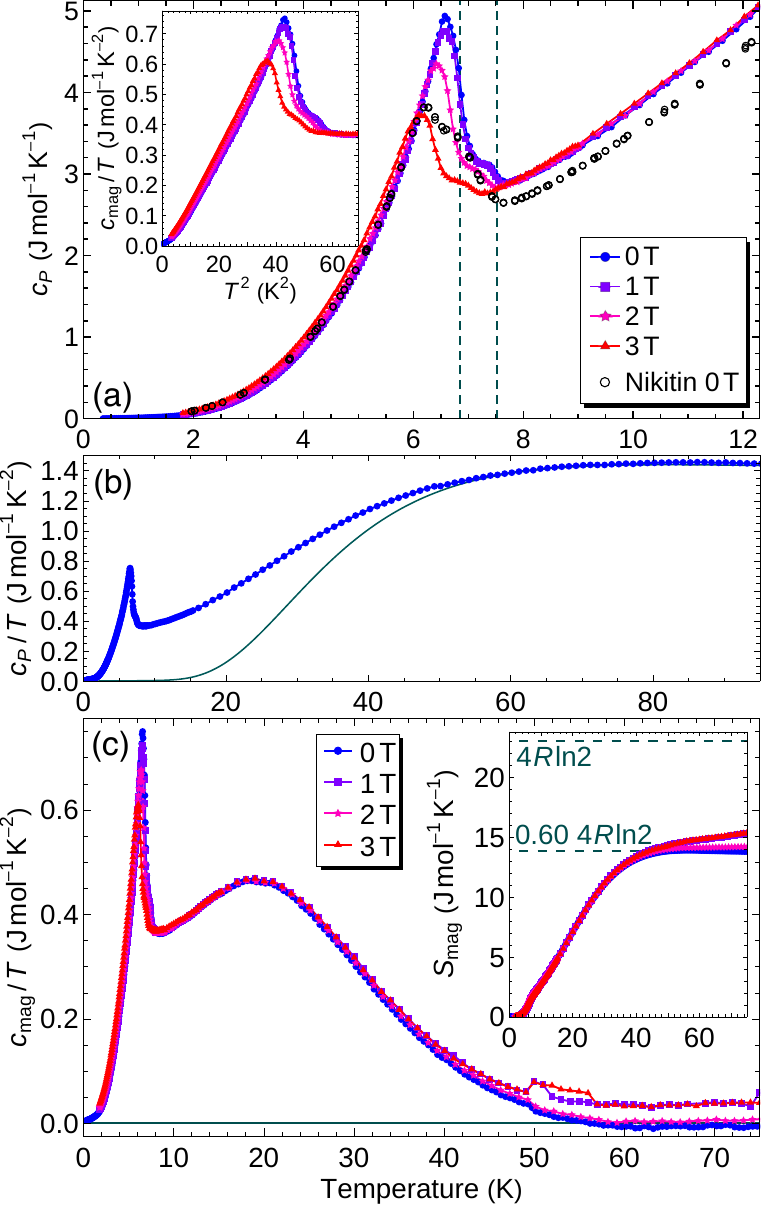}
  \caption{\label{cP}(a) Specific heat of our synthetic brochantite at zero field and for fields along [100].  The inset shows the magnetic component $\cmag/T$ {\itshape vs}.\ $T^2$.  Dashed lines mark the transition temperatures at zero field;  black circles are taken from Ref.~\cite{Nikitin2023}.  (b) $c_P/T$ to higher temperatures, plotted together with the phonon fit. (c) Magnetic component of the specific heat, plotted as \cmag/$T$.  The magnetic entropy is shown in the inset.}
\end{figure}

\begin{figure*}
  \includegraphics[width=\textwidth]{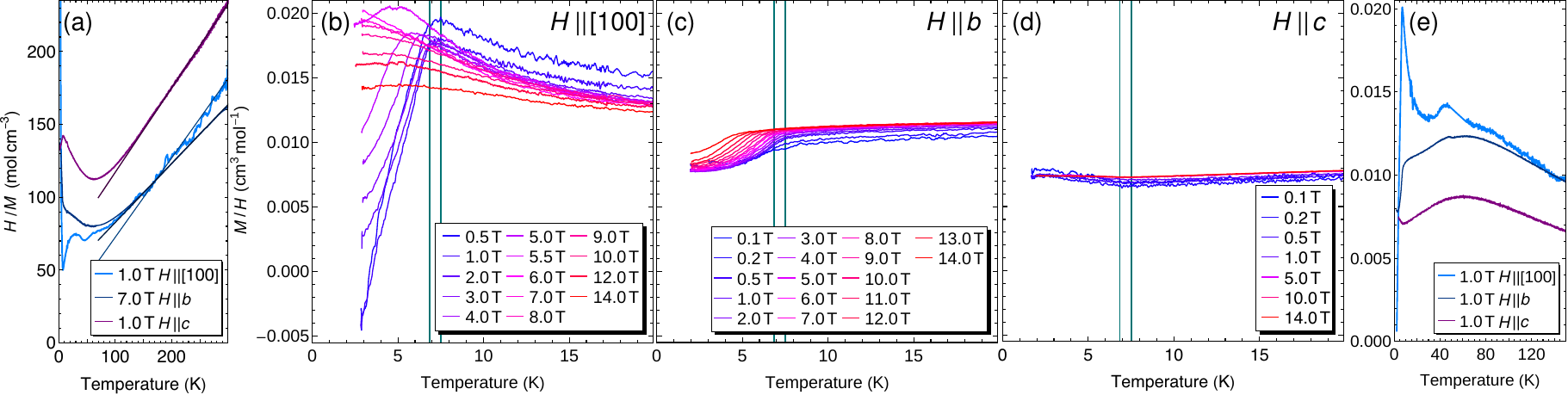}
  \caption{\label{MT}Temperature-dependent magnetization $M/H$ of synthetic brochantite. (a) Curie-Weiss fits to the inverse magnetization.  (b--c) Field-cooled magnetization at low temperature in various fields, for $H\parallel [100]$, $b$, and $c$, respectively, on the same vertical scale.  (d) Magnetization in a 1-T field to higher temperature for all three directions.}
\end{figure*}

The specific heat of our brochantite crystals is shown in Fig.~\ref{cP}(a) for zero field and fields along [100].  On cooling, a weak jump is encountered around 7.5\,K, followed by a stronger jump around 6.85\,K, then no further transitions are obvious down to 0.35\,K.  The overlap of the two transitions makes it difficult to determine their temperatures precisely.  In such a low-dimensional system it would not be surprising to see a broad hump above the main transition, corresponding to short-range order, but the jump at 7.5\,K appears to be narrower than the lower-temperature transition and they are quite close together.  While the transitions are relatively sharp, they are not as sharp as those we have observed in rouaite\,\cite{Aswathi2024} or antlerite\,\cite{Kulbakov2022b} prepared by similar methods in our group, possibly indicative of the higher structural disorder in brochantite.  

Figure~\ref{cP}(a) also includes data from a recent report on natural brochantite by Nikitin {\itshape et al}.\,\cite{Nikitin2023} (black open circles), which is also suggestive of a double transition.  The transitions in the natural sample appear at somewhat lower temperatures, and the lower-temperature peak is less pronounced.  This is surprising given the degree of structural disorder in our crystals.  Natural minerals may be expected to exhibit a higher degree of crystalline perfection due to a slower growth process, possibly at lower temperature;  however, there is also a risk that a significant fraction of the periodic table may be present, which may add additional strong disorder to the magnetic sublattice.  In this regard, the inclusion of Li, Na and Al leaves room for improving our synthesis.  The differences in $c_P$ in the paramagnetic state are likely due to uncertainties in the sample mass;  most likely the data should match at high temperatures and the natural crystal would have a slightly higher specific heat below the transitions.  

Since no nonmagnetic analogue of brochantite has been reported, we extracted the approximate magnetic component of the specific heat \cmag\ by a Debye-Einstein fit to the data above 50\,K: 
\begin{align}
c_{\text{lattice}}(T)~=~&9n_DR \left( \frac{T}{\theta_D} \right)^3 \int_0^{\theta_D/T} \frac{x^4 e^x}{(e^x - 1)^2} \, dx \notag \\
&+ 3n_ER \left( \frac{\theta_E}{T} \right)^2 \frac{e^{\theta_E/T}}{(e^{\theta_E/T} - 1)^2}
\end{align}
Here, $R$, $x$, $n_D=15.3\pm1.6$, $n_E=5.65\pm0.11$, $\theta_D=974\pm19$\,K, and $\theta_E=168.5\pm1.7$\,K represent the ideal gas constant, $\hbar\omega/k_BT$, the Debye coefficient, Einstein coefficient, Debye temperature, and Einstein temperature, respectively.  If the fit is instead performed above 70\,K, the fit parameters change at most by a few parts per thousand, well within their uncertainty.  $n_D$ and $n_E$ sum to 21, the total number of atoms in a formula unit, as expected.  

The phonon component is compared to our zero-field data, plotted to higher temperature as $c_P/T$, in Fig.~\ref{cP}(b).  As in other frustrated and low-dimensional magnetic systems, magnetic entropy associated with the formation of short-range order is visible above the transition, in this case up to $\sim$50\,K, several times \TN.  A similar result was recently reported from a combined fit to the phononic and magnetic specific heat in a natural mineral sample, indicating that magnetic entropy survives to at least 25\,K\cite{Ginga2025}.  The extracted magnetic component of the specific heat, \cmag, is plotted as $\cmag/T$ in Fig.~\ref{cP}(c), where a large apparent hump associated with short-range correlations is centered around 20\,K.  This is distinct from the jump at 7.5\,K, suggesting that the double transition is indeed real.

The magnetic entropy $S_\text{mag}$, extracted from \cmag\ by integrating $\cmag/T$ starting from the lowest-temperature data point, is plotted in the inset to Fig.~\ref{cP}(c).  It asymptotes 60\%\ of the expected $4R\ln2$ for \brochH.  Similar reductions in entropy have also been reported in other frustrated systems \cite{Bag2021,Bhattacharya2024,Singh2024,Kaushick2025Mn} and may result from short-range magnetic correlations persisting to higher temperature, overestimation of the lattice contribution, or a significant release of entropy at temperatures below our minimum temperature of 357\,mK. 

We replot the magnetic component of our specific heat data as $\cmag/T$ vs $T^2$ in the inset to Fig.~\ref{cP}(a).  Below the transitions, this plot is strikingly close to linear over a broad temperature range, indicating a dominant $T^3$ contribution to the specific heat between roughly 2.5 and 6\,K.  The gap seen in inelastic neutron scattering on natural-mineral samples of 1.0(1)\,meV\,\cite{Nikitin2023} would correspond to a temperature well in excess of \TN, and a power law in the specific heat would not be expected to arise from thermally populated excitations across a large gap.  However, the gap was reported based on data collected at 1.7\,K, and may be temperature dependent.  The power-law-like behaviour may arise through a combination of exponential activation across the gap and the temperature dependence of that gap.  

\section{Magnetization}

In our temperature-dependent field-cooled (FC) magnetization data, shown in Figs.~\ref{MT}(b--d), a clear antiferromagnetic-like transition is seen for fields along [100] and $b$, but the magnetization increases slightly below this transition in fields along $c$.  The scale of the changes indicates that the moments lie mainly along [100] and perpendicular to $c$, consistent with the magnetic structure in Ref.~\cite{Nikitin2023}.  The transitions from specific heat are indicated --- the transition for all three directions most likely corresponds to the 6.8-K transition in the specific heat, but due to the noise in the magnetization data this cannot be determined with certainty for $b$ or $c$.  Zero-field-cooled data (not shown) were similar, with a slight upturn visible at the lowest fields for $H\parallel b$ and a slight negative offset.  The most likely explanation for this upturn may be small quantities of impurity phases clinging to the surface of the crystal despite our attempts to clean them before the measurement.  For $H\parallel [100]$ there is also a contribution from the film on which the crystals were mounted which we were not able to subtract, which may lead to a small offset.

\begin{figure*}
  \includegraphics[width=\textwidth]{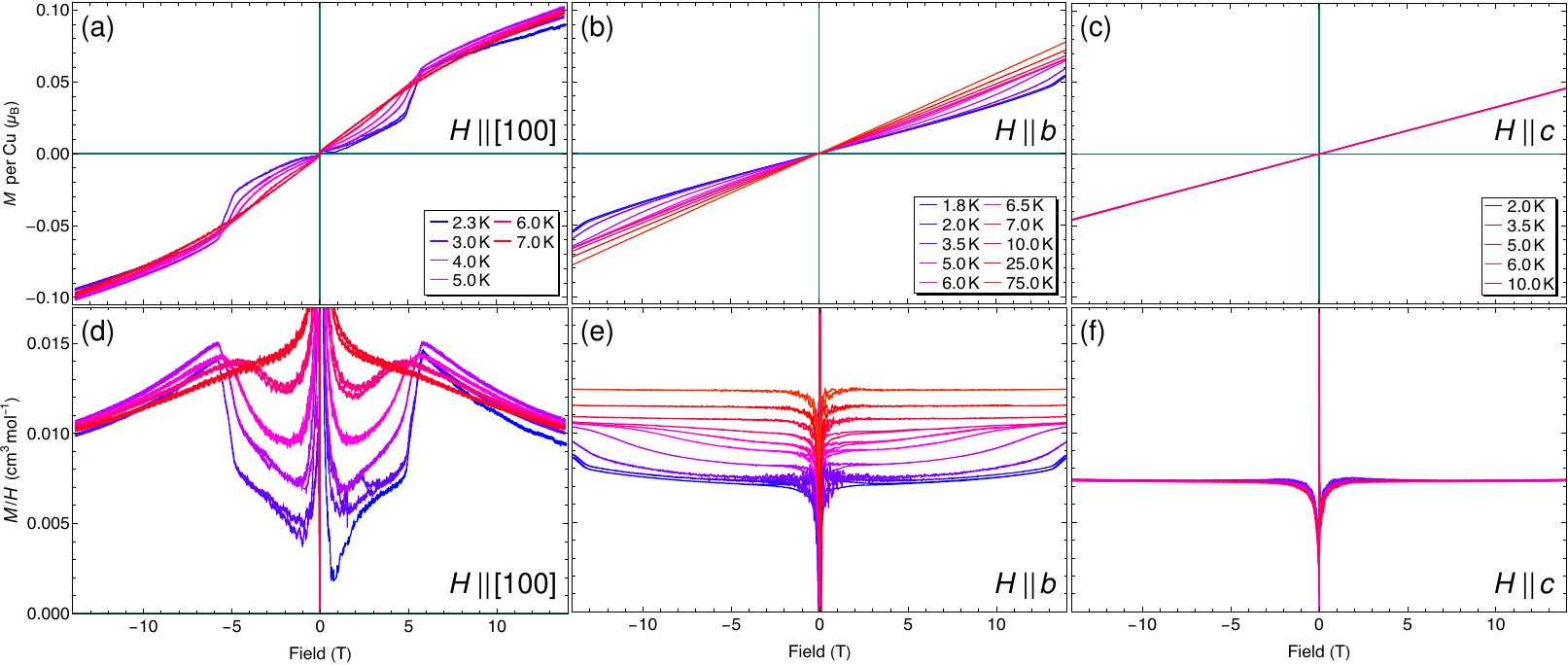}
  \caption{\label{MH}(a--c) Field-dependent magnetization of synthetic brochantite for $H\parallel [100]$, $b$, and $c$, respectively. (d--f) The same data replotted as $M/H$.}
\end{figure*}

Figure~\ref{MT}(e) shows the magnetization to higher temperature, where we observe a broad hump around 60\,K, presumably associated with short-range order within the double chains.  A similar hump was reported in Refs.~\cite{Vilminot2006,Nikitin2023}.  A narrow feature around 50\,K for $H\parallel [100]$ most likely originates from contaminants in the helium exchange gas.

The inverse magnetization is shown in Fig.~\ref{MT}(a).  A Curie-Weiss fit was performed above 170\,K, resulting in a paramagnetic moment of 2.22\,$\mu_\text{B}$/Cu for $H\parallel b$ and a strongly antiferromagnetic Curie-Weiss temperature of $-$104\,K.  Similar fits yielded 2.09\,$\mu_\text{B}$/Cu and $-$82\,K for $H\parallel [100]$ and 1.85\,$\mu_\text{B}$/Cu and $-$101\,K for $H\parallel c$.  The Curie-Weiss temperatures are more than an order of magnitude larger than \TN, indicating strong frustration.

The field-dependent magnetization for fields along [100] and $b$, shown in Figs.~\ref{MH}(a) and \ref{MH}(b) and replotted as $M/H$ in Figs.~\ref{MH}(d) and \ref{MH}(e), respectively, shows a magnetic transition which is initially around 5.5\,T at 2.3\,K for $H\parallel[100]$ and 14--15\,T at 1.8\,K for $H\parallel b$.  These broaden and fall to lower field as the temperature increases, eventually disappearing around \TN.  For $H\parallel [100]$ this may well be \TN, since once it is suppressed it becomes very difficult to identify a \TN\ in Fig.~\ref{MT}.  We note, however, that the lack of saturation indicates that the spin system is far from field-polarized here, so even if long-range order has been destroyed, strong short-range antiferromagnetic correlations must still exist.  Given that a 14-T field for $H\parallel b$ is only able to suppress \TN\ to $\sim$4\,K in $M(T)$ [see Fig.~\ref{MT}(b)], that our $M(H)$ data only reach $\sim$0.1\,$\mu_\text{B}$/Cu at 14\,T [Fig.~\ref{MH}(a--c)], and that previous field-dependent magnetization data showed no signs of saturation up to 30\,T\,\cite{Nikitin2023} for any field direction, this transition is more likely a metamagnetic transition within the ordered phase and not a loss of long-range order.  For fields along $c$, shown in Figs.~\ref{MH}(c,f), the curves are surprisingly temperature-independent.  The only differences among the temperatures measured are weak enhancements in $M/H$ below $\sim$5\,T.  The $M(H)$ curves do not exhibit any significant hysteresis.  

Magnetization data in Ref.~\onlinecite{Nikitin2023} broadly agree with our results.  Based on magnetization and neutron diffraction results, they proposed a magnetic ground state consisting of ferromagnetic chains, with spins on the Cu1--Cu2 chains antialigned with those on the Cu3--Cu4 chains and all spins directed along $\pm a$; however, with detectable magnetic intensity on only two reflections, it was necessary to assume that the spins were collinear and no angles could be determined.  The metamagnetic transition in fields $H\parallel b$ indicates that there must also be a spin component along $b$ which could not be resolved in the previous neutron diffraction work.  This transition can be explained if the reported collinear order at zero field is rotated toward $b$.  However, such a rotation cannot explain the metamagnetic transition for $H\parallel [100]$.

Applying a field of only $\sim$5\,T along [100] runs the system through a metamagnetic transition which roughly doubles the magnetization [Figs.~\ref{MH}(a) and \ref{MH}(d)];  this doubling is consistent with a spin-flop transition, or could represent flipping a [100] component of half the spins.  If brochantite were a collinear antiferromagnet, as previously proposed\,\cite{Nikitin2023}, even assuming a small rotation of the spins toward $b$, flipping half the spins along [100] would bring the system very close to saturation.  However, the same report found no saturation of the magnetization up to 30\,T at 1.5\,K, and our $M(H)$ data in Fig.~\ref{MH} are also very far from saturation at all fields.  In the case of canting we could only be flipping a small component of the order, which would imply at least one additional canting angle and the staggering of some component of the spins along [100] in the ground state, presumably between adjacent ladders or planes.  Neutron diffraction would be required to fully identify the canting and verify whether the transition for $H\parallel[100]$ is a spin flop.  Since no metamagnetic transition is visible up to 30\,T for fields along $c$, the moments most likely lie in the $ab$ plane.

Similar metamagnetic transitions were observed in rouaite\,\cite{Aswathi2024}, where a full description of the ground state allowed tentatively identifying which spin components were flipped at the transitions.  The availability of synthetic brochantite single crystals allows the preparation of deuterated crystals, which will offer an opportunity to fully clarify the magnetic ground state, and thereby understand these transitions, in future works.  

\section{Crystal Structure}

\begin{figure}[b]
  \includegraphics[width=\columnwidth]{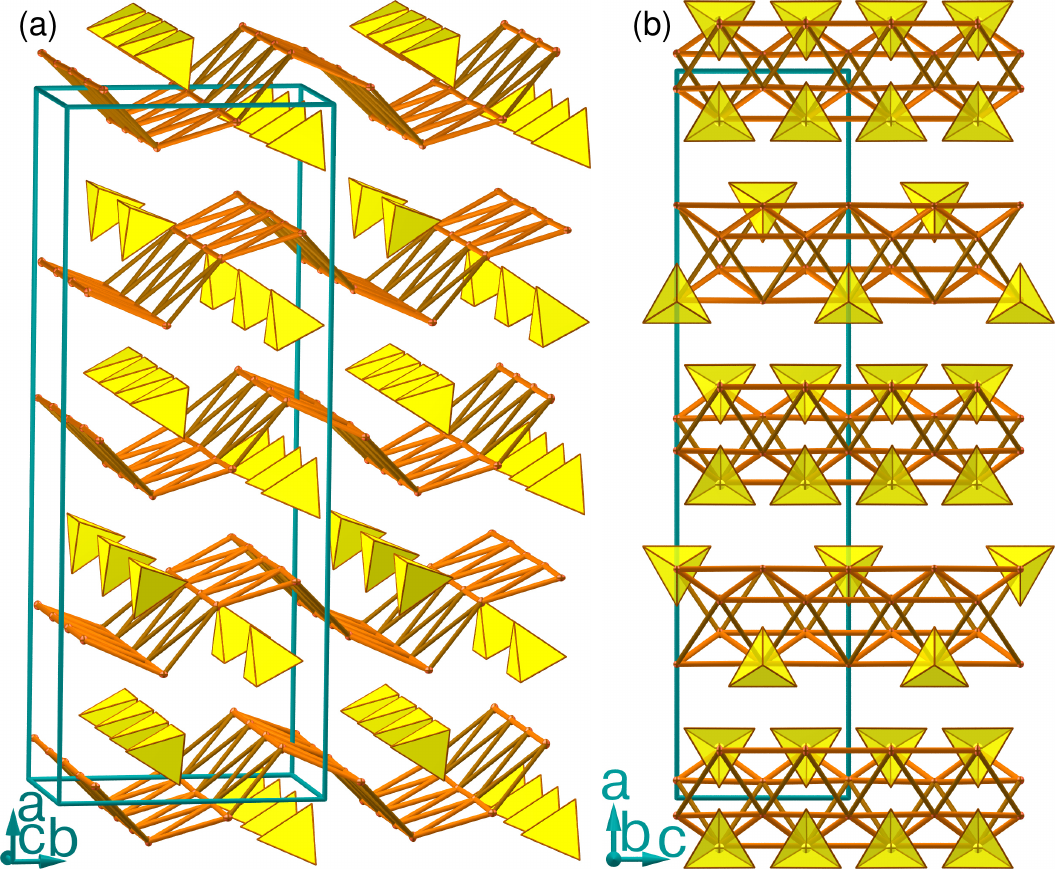}
  \caption{\label{ours}Structure refined from our single-crystal synchrotron diffraction data in $Bb2_1m$.  For clarity, only the Cu planes and SO$_4^{2-}$ tetrahedra are shown.  The layers at the top, bottom, and middle of the unit cell have twice as many SO$_4^{2-}$ tetrahedra because these are roughly half occupied, which is likely an artefact of stacking disorder.  The remaining two layers differ primarily in a shift along $c$ of the surrounding SO$_4^{2-}$ anions.}
\end{figure}

\begin{figure}[tb]
    \includegraphics[width=\columnwidth]{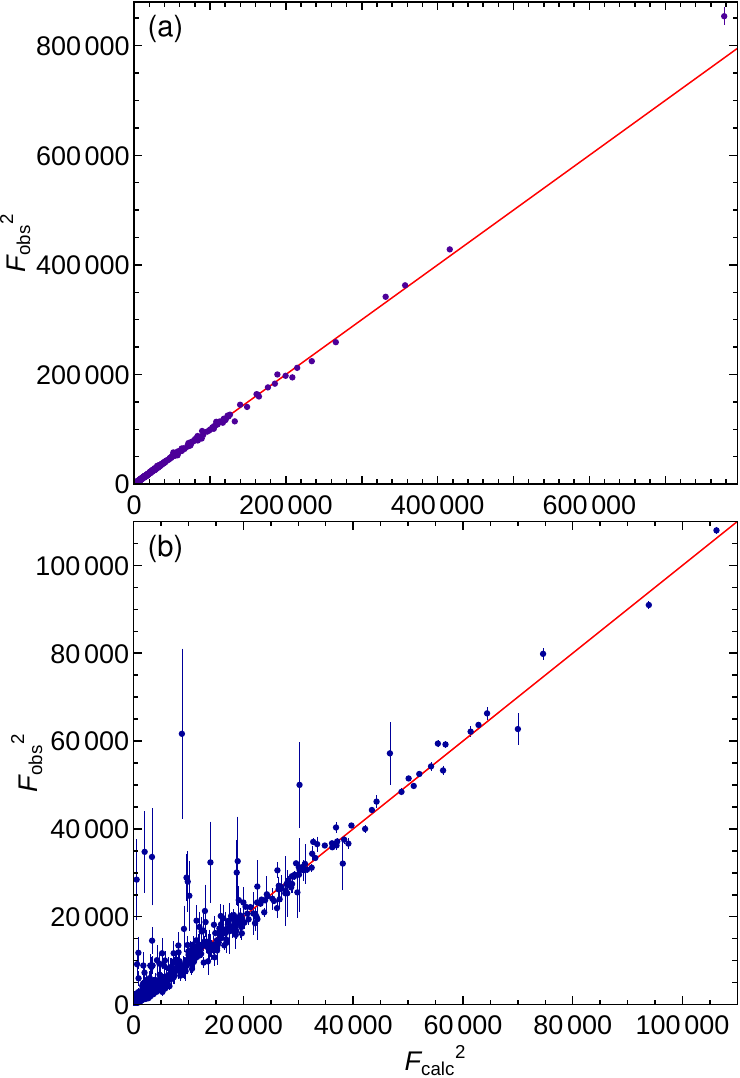}
    \caption{\label{FvsF}$F_\text{calc}^2$ {\itshape vs}.\ $F_\text{meas}^2$ characterizing our structure refinements at room temperature (a) with x-rays at BM01, ESRF, and (b) using the Koala neutron Laue diffractometer.}
\end{figure}

Using a laboratory single-crystal diffractometer, it was not straightforward to distinguish the structure of brochantite from a doubled, orthorhombic cell similar to that reported for discredited orthobrochantite, with $a$=25.52\,\AA, $b$=9.86\,\AA, and $c$=6.03\,\AA.  We clarified this with single-crystal synchrotron diffraction at beamline BM01 (Swiss-Norwegian Beamline) at the ESRF in Grenoble, France, leading to the refinement summarized in Table~\ref{refineTab1} and characterized by the plot of $F^2_\text{calc}$ {\itshape vs}.\ $F_\text{meas}^2$ in Fig.~\ref{FvsF}(a).  Additional powder diffraction data were collected on beamline BM01 to check the most appropriate space group as described in Appendix~\ref{AppA}.  Our refinements found planes with double the usual number of SO$_4^{2-}$ sites, roughly half filled. This added up to a roughly 1.5\% SO$_4^{2-}$ excess, less than the $\sim$20\% S excess from EDX.  This discrepancy is readily explained by the limited sensitivity of EDX to light elements.  

\begin{figure}[t]
    \includegraphics[width=\columnwidth]{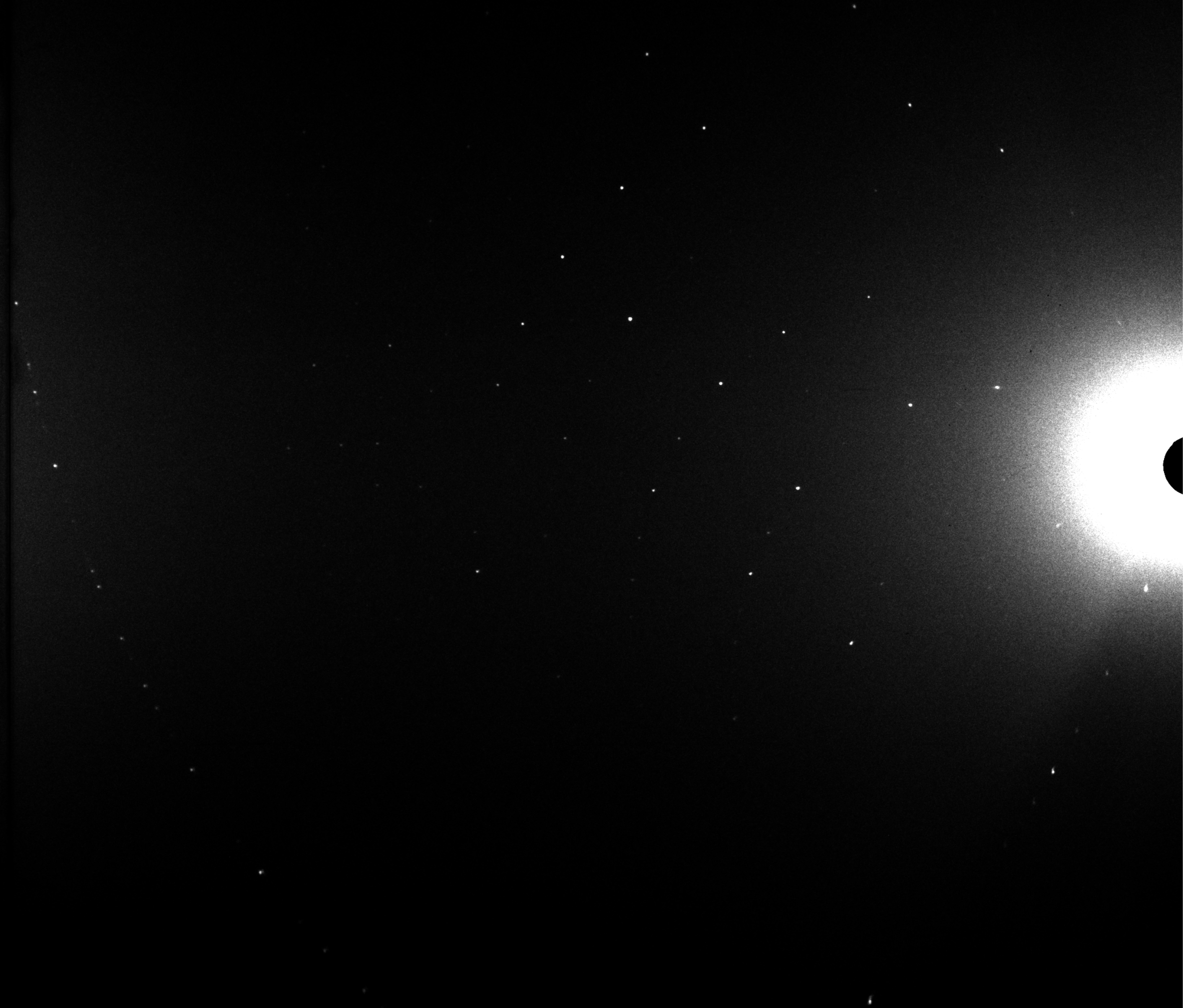}
    \caption{\label{Koala}Portion of a neutron Laue pattern of one of our crystals at room temperature.}
\end{figure}

Our refined crystal structure is shown in Fig.~\ref{ours}.  It exhibits ABAC layer stacking, where the A layers have twice as many SO$_4^{2-}$ groups, all roughly half occupied.  The stacking from a B layer to a C layer is suggestive of MDO$_1$ stacking, but with considerable twinning.  

The SO$_4^{2-}$ vacancies were investigated by neutron Laue diffraction on the Koala diffractometer at ANSTO, Sydney, Australia.  An example frame is shown in Fig.~\ref{Koala}, which also serves to demonstrate the crystal quality.  Our refinement found that the SO$_4^{2-}$ vacancies are occupied by two (OH)$^-$ groups, and returned a Cu:S ratio of 4:0.97(16).  Details of this refinement are provided in Table~\ref{refineTab1} in Appendix~\ref{AppA}, and a plot of $F_\text{calc}^2$ {\itshape vs}.\ $F_\text{meas}^2$ to characterize the quality of this structure refinement is shown in Fig.~\ref{FvsF}(b).  Since the refined compositions from both x-ray and neutron diffraction are close to that of stoichiometric brochantite, this doubling of sulphate sites in some layers with half occupation presumably indicates disordered stacking of what we assume to be ideal brochantite layers.  The A, B, and C layers are presumably identical, but flipped along [100] or shifted along $b$ relative to each other.  We next proceed to investigate the stacking disorder through diffuse x-ray scattering.

\begin{figure}[t]
  \includegraphics[width=\columnwidth]{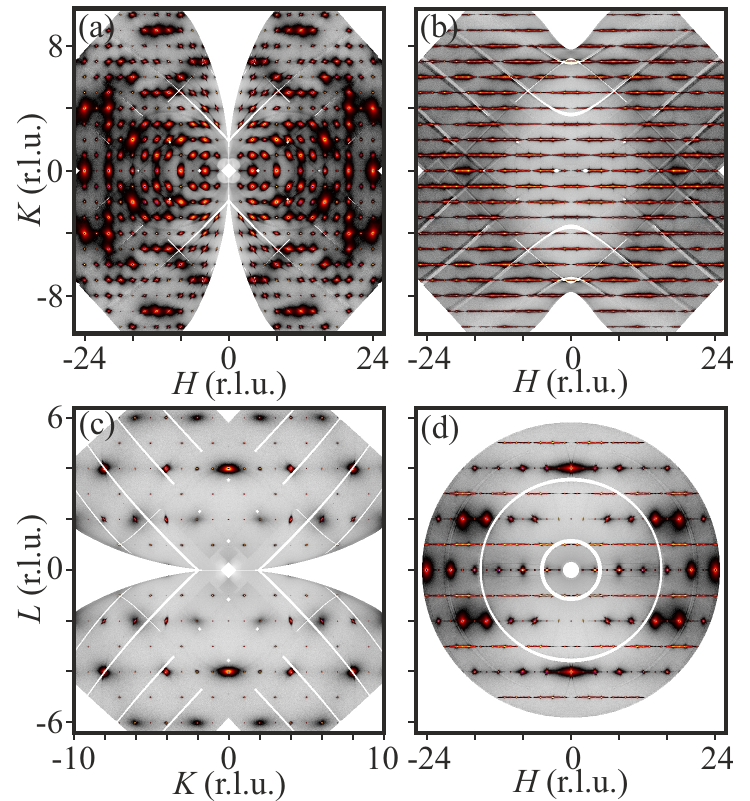}
  \caption{\label{diffuse_slices}Slices through the single-crystal synchrotron diffuse scattering pattern, in the (a) $(HK0)$, (b) $(HK1)$, (c) $(0KL)$ and (d) $(H0L)$ planes in $Bb2_1m$.}
\end{figure}

Slices through our single-crystal diffuse synchrotron diffraction data indexed assuming a doubled, orthorhombic cell are shown in Fig.~\ref{diffuse_slices}.  As can be seen, sharp peaks appear at $(HKL)$ positions for odd $H$, supporting the presence of the longer repeat unit, but these spots are sitting atop diffuse rods of scattering which run along $\text{\itshape H}$.  This is indicative of stacking faults.  The doubled, orthorhombic structure is ABAC stacked, where B and C are related by a translation of $\frac{c}{2}$.  A strict adherence to ABAC stacking would make the $a$ lattice parameter 25\,\AA, but stacking faults lead to ABAB or ACAC stacking locally, halving the unit cell locally and reducing the intensity of the odd-$H$ reflections.  

\begin{figure}[tb]
  \includegraphics[width=\columnwidth]{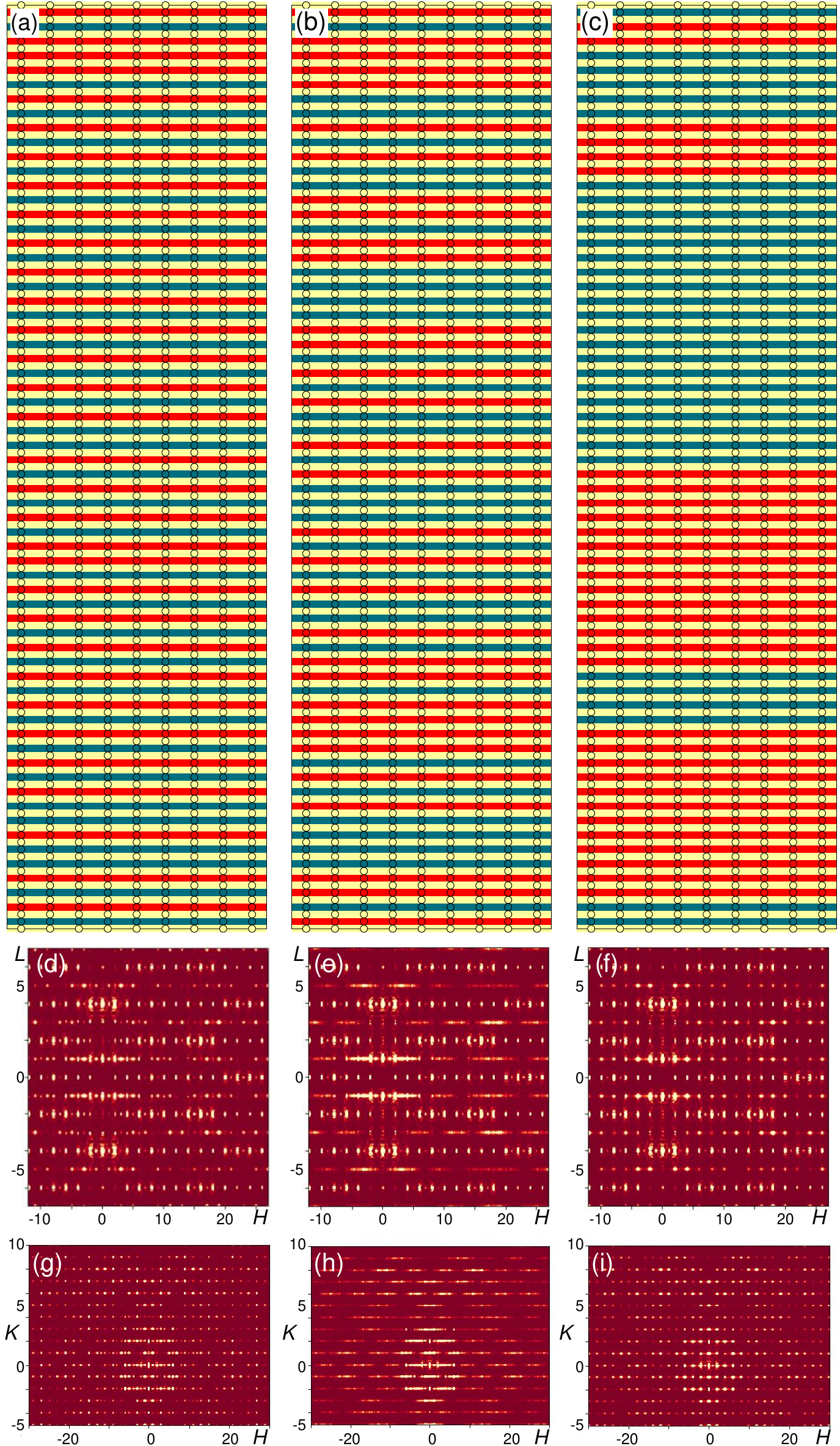}
  \caption{\label{Monte}Monte-Carlo stacking patterns for layer correlations (a) $-0.81$, (b) +0.11, and (c) +0.87; in the ABAC stacking, A corresponds to yellow circles, B is blue, and C is red.  These would produce the corresponding scattering sections along (d-f) $(H0L)$ and (g-i) $(HK1)$.}
\end{figure}

We made a structural model with alternating layers of A with either B or C, and used a Monte Carlo algorithm to order them according to a specified correlation ($-$1 being pure ABAC and +1 being clusters of like layers --- ABAB or ACAC).  Results depicting three different levels of correlation are shown in Figs.~\ref{Monte}(a-c), where yellow, red, and blue correspond to layers A, B, and C, respectively.  The resulting simulated $(HK0)$ and $(1KL)$ scattering planes are shown in Figs.~\ref{Monte}(d-f) and \ref{Monte}(g-i), respectively.  A comparison to Fig.~\ref{diffuse_slices} suggests nearly random layer stacking.  

\section{Conclusion}

The growth of single crystals of brochantite offers a synthetic starting point for exploring this material and its uniquely rippled distorted-triangular Cu$^{2+}$ planes in more detail.  Our crystals have nearly random layer stacking, but exhibit a weak tendency toward ABAC stacking, which would be a new polytype; this would presumably also exist in natural mineral form.  Despite their considerable stacking disorder and potential anion disorder, our crystals have sharper transitions than previously investigated natural mineral samples, suggesting that purity on the Cu site may be more important than other forms of disorder.  It is likely that tuning the synthesis conditions, particularly temperature, would have an impact on the stacking disorder, opening the possibility to investigate the effect of stacking disorder on the physical properties, in particular the magnetic order.  Since the copper layers are linked to each other only through hydrogen bonds, the interlayer interactions are presumably weak and the stacking disorder is likely to have only a minor impact, but this can now be investigated.  The previously proposed magnetic structure is evidently incomplete, and we suggest how it would need to be modified; the existence of clean single-crystalline samples will enable a more complete magnetic structure refinement.  

\section*{Data Availability}

Samples and data are available upon reasonable request from D.\ C.\ Peets or D.\ S.\ Inosov; data underpinning this work is available from Ref.~\onlinecite{OPARA_Needles}.

\begin{acknowledgments}
The authors are grateful to S.V.\ Krivovichev for helpful discussions and S.\ Nikitin for illuminating discussions and for sharing specific heat data.  This project was funded by the Deutsche Forschungsgemeinschaft (DFG, German Research Foundation) through:  individual grant PE~3318/2-1 (Project No.\ 452541981); projects B03, C01, C03, and C10 of the Collaborative Research Center SFB~1143 (Project No.\ 247310070); Research Training Group GRK~1621 (Project No.\ 129760637); and the W\"urzburg-Dresden Cluster of Excellence on Complexity and Topology in Quantum Materials\,---\,\textit{ct.qmat} (EXC~2147, Project No.\ 390858490). The PPMS Dynacool-12 at TUBAF was funded through DFG Project No.\ 422219907.  We acknowledge the ESRF for provision of synchrotron radiation facilities under proposal HC-4948.  The authors acknowledge the support of the Australian Centre for Neutron Scattering, Australian Nuclear Science and Technology Organisation, in providing the neutron research facilities used in this work. 
\end{acknowledgments}

\appendix
\section{Crystal Structure Refinement Details}\label{AppA}

\begin{figure}[tb]
  \includegraphics[width=\columnwidth]{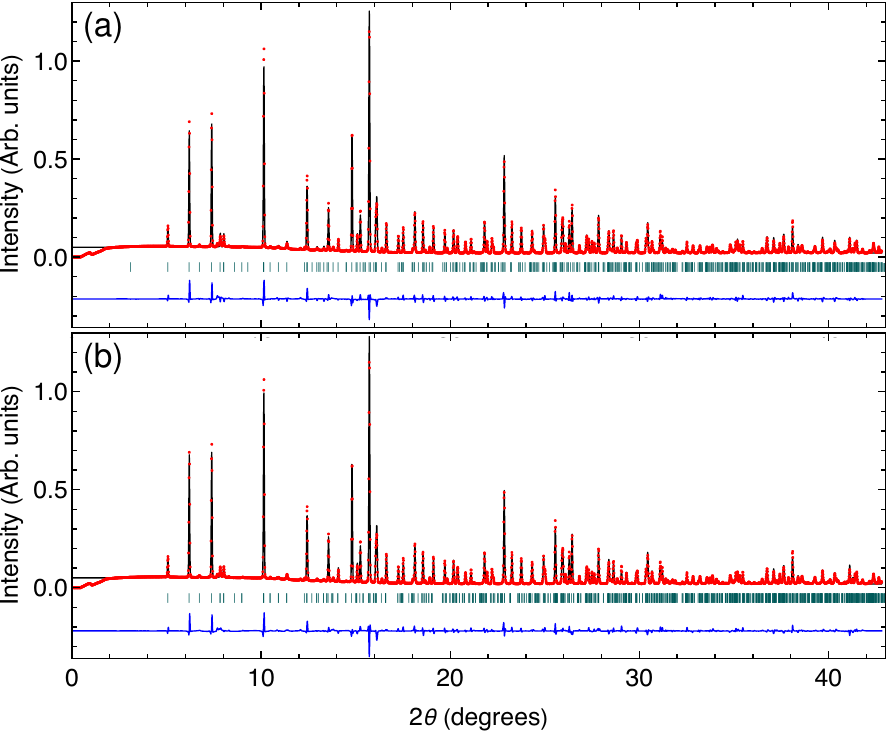}
  \caption{\label{BM01}Crystal structure refinements in (a) $Bb2_1m$ and (b) $P2_1/n$ of our synchrotron powder diffraction data collected at beamline BM01 at the ESRF.}
\end{figure}

Additional powder diffraction data were collected on beamline BM01 to check the most appropriate space group;  refinements are shown in space groups $Bb2_1m$ --- an alternative setting of $Cmc2_1$, \#\ 36, chosen to match literature reports --- and $P2_1/n$ (\#\ 14) in Figs.~\ref{BM01}(a) and \ref{BM01}(b), respectively.  The axes in $P2_1/n$ are permuted relative to $Bb2_1m$ and the long axis was not doubled to match orthobrochantite as in the other refinements.  These correspond to MDO$_1$ and MDO$_2$, respectively.  The fit was marginally better in the former.  

Key parameters and results of our crystal structure refinements are summarized in Tab.~\ref{refineTab1}.  Due to the large number of atoms in the unit cell, we do not reproduce atomic positions, anisotropic displacement parameters, bond lengths, or bond angles here.  These are available in CIF files provided as arXiv ancillary files.  The quality of the refinement is characterized through the plot of $F^2_\text{calc}$ {\itshape vs}.\ $F_\text{meas}^2$ in Fig.~\ref{FvsF}(a).

\begin{table}[b]
  \caption{\label{refineTab1}Details of our crystal structure refinements of synthetic brochantite based on single-crystal diffraction data taken at beamline BM01 at the ESRF and Laue diffraction data taken on the Koala diffractometer at ANSTO, Australia.  Note that neutron Laue diffraction is relatively insensitive to absolute lattice parameters. Only reflections with intensity $I>3\sigma$ were considered for the Koala refinement. For further details, e.g.\ atomic positions, please refer to the CIFs provided as arXiv ancillary files.}
  \begin{tabular}{lll}\hline\hline
  Parameter & BM01 & Koala \\ \hline
  Temperature (K) & 293(2) & 293(2)\\
  Wavelength & 0.65524\,\AA & White beam\\
  Space group & $Bb2_1m$ (\#\ 36) & $Bb2_1m$ (\#\ 36)\\
  $a$ (\AA) & 25.5231(7) & 25.55\\
  $b$ (\AA) & 9.8620(2) & 9.85\\
  $c$ (\AA) & 6.0300(2) & 6.03\\
  $V$ (\AA$^3$) & 1517.80(7) & 1517.56\\
  $Z$ & 8 & 8 \\
  Density (g/cm$^3$) & 3.936 & 3.95\\
  Reflections & 4170 & 2549\\
  Reflections ($I>2\sigma$) & 3109 & \\
  Reflections ($I>3\sigma$) &  & 596\\
  $hkl$ range probed & $-36\leq h\leq 44$ & $0\leq h\leq 44$\\
   & $-17\leq k\leq 14$ & $0\leq k\leq 19$\\
   & $-10\leq l\leq 7$ & $0\leq l\leq 12$\\
  F(000) & 1722 & 560.258\\
  $R$ (all reflections) & 4.80\% & 8.95\%\\
  $R$ ($I>2\sigma$) & 2.92\% & ---\\
  $wR$ (all reflections) & 7.05\% & 9.89\%\\
  $wR$ ($I>2\sigma$) & 6.42\% & ---\\ 
  Flack parameter & 0.41(2) & ---\\ \hline\hline
  \end{tabular}
\end{table}

\bibliography{needles}

\section*{Supplemental Material: Details of structure refinements}

Due to the large number of atoms in an orthorhombic unit cell, we do not reproduce tables of the atomic positions, anisotropic displacement parameters, bond lengths, or bond angles here.  Crystallographic information files (CIFs) are provided instead, as arXiv ancillary files.  In all cases, the CIFs are provided both (1) in their original form in the conventional $Cmc2_1$ setting of space group \#~36, and (2) in the $Bb2_1m$ setting to ease comparison to literature refinements of brochantite.  The files available are summarized below:

\begin{table}[htb]
  \vspace{8pt}\noindent\begin{tabular}{lcr}\hline\hline
    Filename & Source & Setting\\ \hline
    \verb+BM01_AC-6A_Cmc21.cif+ & BM01, ESRF & $Cmc2_1$\\
    \verb+BM01_AC-6A_Bb21m.cif+ & BM01, ESRF & $Bb2_1m$\\
    \verb+Koala_293K_AC-6A_Cmc21.cif+ & Koala, ANSTO & $Cmc2_1$\\
    \verb+Koala_293K_AC-6A_Bb21m.cif+ & Koala, ANSTO & $Bb2_1m$\\ \hline\hline
  \end{tabular}
\end{table}
\end{document}